# Observation of Weak Localization of Seismic Waves

E. Larose[1], L. Margerin[1*], B.A. van Tiggelen[2], M. Campillo[1,2]
[1]*Laboratoire de Géophysique Interne et Tectonophysique, Observatoire de Grenoble,
Université Joseph Fourier& CNRS, BP 53, 38041 Grenoble, France.*
[2]*Laboratoire de Physique et Modélisation des Milieux Condensés,
Université Joseph Fourier & CNRS, Maison des Magistères, BP 166, 38042 Grenoble, France.*

We report the observation of weak localization of seismic waves in a natural environment. It emerges as a doubling of the seismic energy around the source within a spot of width a wavelength, that is several tens of meters in our case. The characteristic time for its onset is the scattering mean free time, that quantifies the internal heterogeneity.

Weak localization (WL) is a manifestation of interference of multiply scattered waves in disordered media. It was first introduced 20 years ago in quantum physics to explain novel features in the electronic magneto-resistance at low temperatures [1–3], and initiated a genuine explosion of mesoscopic physics. The discovery of WL constituted the desired counter-example of the one-century old assertion that multiple scattering of waves destroys wave-phenomena, reducing it conveniently to classical radiative transfer, where waves are treated like hard spheres colliding with obstacles. In optics [4–6] and in acoustics [7] the effect is better known as coherent backscattering, where it was shown to be an accurate way to measure transport mean free paths or diffusion constants. This feature finds its origin in the constructive interference between long reciprocal paths in wave scattering [8, 9]. This enhances the probability to return to the source by a factor of exactly two, that results in the local energy density enhancement by the same factor. In seismic experiments we expect WL to appear as an enhancement of seismic energy in the vicinity of a source [10, 11].

In the heterogeneous Earth the wave propagation becomes complex and wave scattering results in a "seismic coda" [12], which forms the tail of the seismograms. The coda is not always processed, because it is believed not to contain any structural information that is easily extractable using standard imaging techniques. Nevertheless, coda energy decay is widely recognized to be sensitive to the regional geological environment. During the last two decades, radiative transfer was successfully introduced to model the energy decay of coda waves [12]. It describes the transport of the wave energy in space and time, but does not take into account phase information. Radiative transfer predicts the equipartition of waves among different modes [13] which has been observed [14], leading to new approaches for processing coda waves [15, 16]. However, the WL effect has never been observed in seismology. The aim of this work is to show the relevance of mesoscopic physics to seismology and its necessity to interpret observed seismic records. In this paper we present the first observation of WL of seismic waves.

The seismic experiments were undertaken at the *Puy des Goules* volcano (central France). Volcanoes are known to be very heterogeneous and might guarantee multiple scattering [17]. A sketch of the experimental set-up is displayed in Figure 1. We have measured the vertical ground motion using a linear array of 23 geophones. The ground motion is the result of a sledgehammer strike at time $t = 0$ on a 20 cm×20 cm aluminum plate which was repeated 50 times for each location. The impact is a vertical, reproducible force in the 15 Hz − 30 Hz frequency range, and can be considered as a narrow impulse in time and space. Because the receivers are placed at the free surface, the detected waves are both bulk waves (with either compressional or transverse polarization) and surface waves (Rayleigh waves with elliptical polarization), each propagating at its own velocity. The wavelengths $\lambda$ are roughly ranging from 9 m (30 Hz Rayleigh waves) to 40 m (15 Hz compressional waves). A typical record is presented in Figure 2.

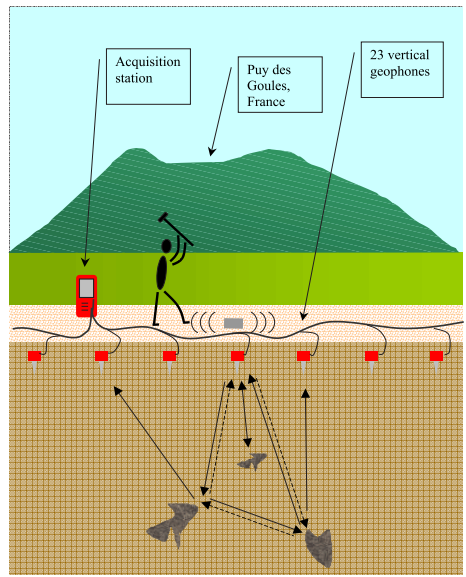

FIG. 1: Experimental setup. Solid and dashed arrows illustrate reciprocal scattered wave paths.

---

[*]To whom correspondence should be addressed; E-mail: Ludovic.Margerin@ujf-grenoble.fr

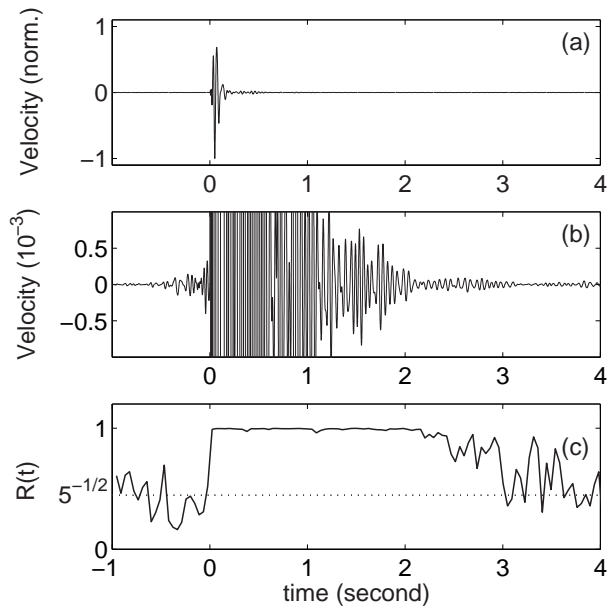

FIG. 2: (a) Example of vertical ground motion signal $s_i$ at the source location. (b) Zoom into the coda. (c) The cumulative ratio $R$ as a function of time, calculated from Eq. 1. $R \approx \frac{1}{\sqrt{5}}$ indicates that the record is dominated by random noise, whereas $R \approx 1$ indicates that the record is strongly dominated by deterministic waves produced by the impact.

The first 0.5 s of the 3 s signal is composed of direct and simply reflected waves, which are traditionally used in seismic prospecting. In this work we will process the average energy of the subsequent seismic coda. The identification of WL must be accompanied by a close study of different kinds of noise that contaminate the seismic record. In the following discussion we separate the ambient noise from the one generated by the operator of the hammer, and identify the mesoscopic regime where noise is negligible. Since ambient noise is generated by meteorological phenomena (like wind) and human activity, the experiments were conducted at night and under anticyclonic conditions. This background noise is stationary and random. All geophones were buried at 20 cm depth to reduce the acoustic signal transmitted by the air and to improve their coupling with the ground. The operator noise is coming from the person manipulating the hammer who is subject to residual movements just before and after the hammer strike. This noise is difficult to separate unambiguously from the signal, because it is local and non-stationary and could be misinterpreted as WL. Fortunately, biophysical studies have revealed that the reproducibility of human motion is limited to frequencies lower than 10 Hz [18]. This suggests that the noise produced by the operator can be considered as random in our frequency band.

We study the sum of $M$ signals $s_i(t)$ produced by repeated strikes at the same location. Each signal results from $N = 10$ strikes that were automatically stacked in the field. We expect both the ambient and the human noise to add up incoherently ($\propto \sqrt{M}$) while the seismic signal deterministically generated by the impacts should add up coherently ($\propto M$). We analyze the time-evolution of the signal-to-noise ratio using the cumulative index $R$,

$$R(t) = \sqrt{\frac{1}{M} \frac{\left\langle \left[\sum_{i=1}^{M} s_i(t)\right]^2 \right\rangle}{\left\langle \sum_{i=1}^{M} s_i^2(t) \right\rangle}}. \quad (1)$$

The brackets denote an average over one oscillation period $T \approx 40$ ms. The ratio $R(t)$ takes its maximum value 1 for a perfectly deterministic signal and equals $1/\sqrt{M}$ for pure random noise. Figure 2-c shows an example of $R(t)$, computed for $M = 5$ signals recorded at the source position. It confirms the randomness of the operator noise ($t < 0$) and the deterministic nature of the seismic signal. Between 0 and 2 s, $R(t)$ always exceeds 90% which enables the processing of the coda with excellent signal-to-noise ratio.

The WL effect finds its origin in the interference of reciprocal, multiply scattered waves, that leads to an enhancement of ensemble-averaged energy of exactly two at the source. Its observation requires the fulfillment of four conditions. Some receivers must be placed less than one wavelength from the source (interference condition). Given the *vertical* force as a source, we must study the energy $E(t)$ associated with the *vertical* seismic motion as a function of source-receiver distance (reciprocity condition) [11, 19]. Thirdly, waves must have the time to scatter at least twice (multiple scattering condition). Finally, enhancement is expected to occur only for the *ensemble-averaged* energy because speckles, i.e. random interference patterns, dominate in a single profile. Because of its random nature the speckle is suppressed by a configurational average while the deterministic WL effect survives. The only average conceivable in seismology is one over source and receiver positions for a *fixed* source-receiver distance $\Delta \mathbf{r}$. For a diffuse field the correlation length is $\lambda/2$ [20], which implies that a different configuration is probed when the whole set-up is moved over more than one wavelength, thus providing us with 12 independent source-receiver configurations.

To evaluate the spatial enhancement of energy $S(\Delta \mathbf{r})$ we normalize the average energy $\langle E_C \rangle$ around the source by its measured average value $\langle E_D \rangle$ sufficiently far away

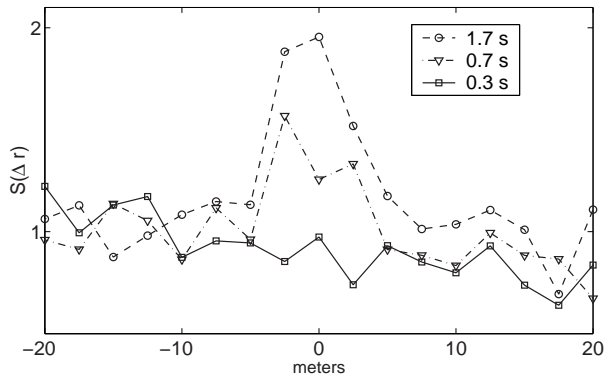

FIG. 3: Energy ratio $S(\Delta\mathbf{r})$ around 20 Hz as a function of source-receiver distance $\Delta\mathbf{r}$ for three different lapse times. The WL effect sets in at a time of roughly 0.7 s, and is fully stabilized at 1.7 s.

(15 m) from the source where the energy density is independent of the source-receiver distance $\Delta\mathbf{r}$. The theoretical prediction for $S(\Delta\mathbf{r})$ at the free surface of an elastic body was obtained in [19]. Since Rayleigh waves likely dominate at the surface, this rigorous expression can be approximated by the profile predicted for 2D random media [21],

$$S(\Delta\mathbf{r}) \equiv \frac{\langle E_C \rangle}{\langle E_D \rangle} \approx 1 + J_0^2\left(2\pi\Delta\mathbf{r}/\lambda\right), \quad (2)$$

where $\lambda$ is the wavelength of the dominating Rayleigh waves and $J_0$ the Bessel function. Note that for the near-field regime the size of the WL spot is independent of lapse time, contrary to the far field regime [7]. The energy distribution $E(t)$ at each sensor is integrated over one sliding window of one cycle duration. The dynamics is studied by analyzing the signals in non-overlapping time windows of 0.4 s duration. In each window, $E(t)$ is normalized at each time $t$ by the maximum over the array, and then averaged over the 12 configurations with equal $\Delta\mathbf{r}$. This procedure compensates for the exponential decay of the total energy, and provides an unbiased average over the different strikes. Finally we integrate the normalized, averaged energy $\langle E(t, \Delta\mathbf{r}) \rangle$ over the entire time window. $S$ is then computed from Eq. (2).

In Figure 3 we plot the seismic energy around 20 Hz measured in the coda as a function of source-sensor distance, and for three specific 0.4 s windows. Around 0.3 s only simply reflected waves are recorded and no energy enhancement is visible around the source. The remaining fluctuations are ascribed to the incomplete suppression of speckle. As from 0.7 s, WL is observed with a gradually increasing enhancement factor at the source. After 1.7 s the profile including the enhancement factor 2 has stabilized, as predicted by the theory for WL in the near field. Therefore, we attribute this enhancement to WL.

According to Eq. (2) the spot has a spatial extent equal to the wavelength $\lambda$. This gives the estimate $c = 260$ m/s for the phase velocity of the Rayleigh waves around 20 Hz. Since at least two scattering events are necessary to generate the enhancement effect, the rise of the enhancement factor corresponds to the transition from the simple to the multiple scattering regime. It was verified in numerical studies [10] that the characteristic time governing the rise of the enhancement factor is the scattering mean free time $\tau$. We thus conclude that this important time scale is of the order of 0.7 $s$ around 20 Hz. For a velocity $c = 300$ m/s this implies a scattering mean free path $\ell \approx 200$ m. We emphasize that this parameter is very difficult to measure with traditional techniques based on attenuation studies because absorption is hard to separate from scattering effects.

We have finally studied the frequency dependence of WL. To this end, the seismograms were filtered in 3 consecutive frequency bands, and the energy profiles were computed as above, though now averaged over the entire coda that exhibits the stabilization of the enhancement $S$ (Figure 4). Three different WL widths are observed. The values for the wavelengths estimated from a fit to Eq. (2) have been indicated. We have separately measured the wavelength of Rayleigh waves from a dispersion analysis of direct arrivals in the original records. Both estimates of the wavelength are consistent and indicate a significant dispersion due to the depth dependence of elastic properties. As a result, the spatial width of WL depends non trivially upon frequency. Future studies might even reveal the frequency dependence of the scattering mean free path $\ell$, which would provide precious information on the nature of the heterogeneity.

In conclusion, we have observed weak localization of seismic waves in a shallow volcanic structure, both in space and time. The observation is in good agreement with the near-field theory for weak localization, which predicts a size of one wavelength for the enhancement spot. The study of this effect turns out to offer a unique opportunity to measure the scattering mean free time *without* the bias of absorption. We found an estimate of 200 m for the mean free path for seismic waves around 20 Hz. Though relatively easy to set-up, our experiment reveals the mesoscopic nature of seismic waves that have travelled many kilometers. As has been the case in nanophysics and in colloid physics, mesoscopic physics may open up new fields of investigation and application in seismology.

We acknowledge support from the French Ministry of Research (ACI *Mésoscopie des Ondes Sismiques*, ACI *Prvention des Catastrophes Naturelles*), GDR IMCODE and program *Intérieur de la Terre* of CNRS/INSU. We wish to thank F. Brenguier, Ch. Voisin, O. Coutant and M. Dietrich for their assistance in the experiment. We thank A. Derode, M. Fink, R. Maynard, A. Paul, J. de

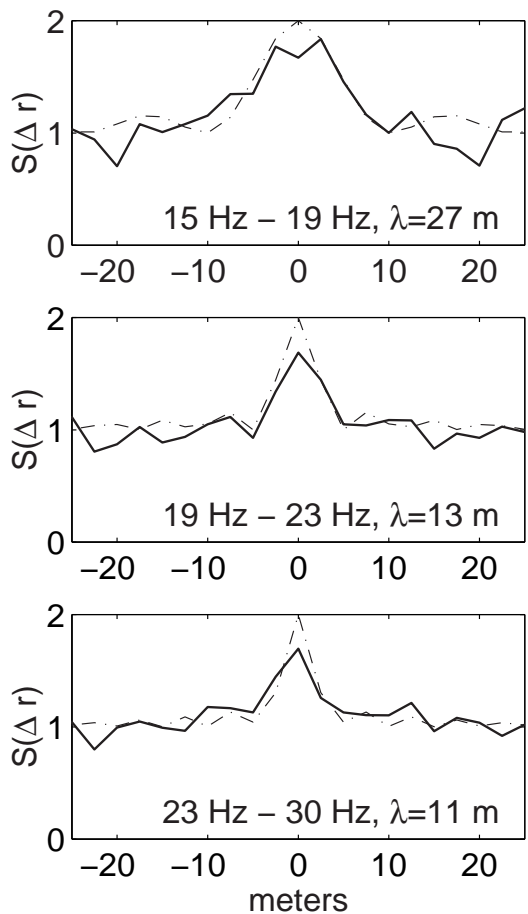

FIG. 4: Observed energy ratio $S(\Delta \mathbf{r})$ (solid lines) for three different frequency bands. The dash-dotted lines represent the theoretical prediction (Eq. 2) fitted for the wavelength $\lambda$.

Rosny, A. Tourin for numerous discussions.